\documentclass[a4paper,fleqn]{cas-sc}
\usepackage{hyperref}

\usepackage[round]{natbib}
\usepackage{graphicx}
\usepackage{xspace}	
\usepackage{graphicx}
\usepackage{lscape}
\usepackage{threeparttable}
\usepackage{amsmath}
\usepackage[utf8]{inputenc}
\usepackage{longtable}
\usepackage{graphicx}
\usepackage{multirow}
\usepackage{rotating}

\newcommand{\ep}{$E_\textrm{p}$}
\newcommand{\tp}{$t_\textrm{p}$}

\newcommand{\rp}{$R_\textrm{p}$}
\newcommand{\lp}{$L_\textrm{p}$}
\newcommand{\vexp}{$V_\textrm{exp}$}
\newcommand{\td}{$t_\textrm{d}$}
\newcommand{\eexp}{$E_\textrm{exp}$}
\newcommand{\mej}{$M_{\textrm{ej}}$}
\newcommand{\spin}{$P_\textrm{i}$}
\newcommand{\trhalf}{$t_{r_{L_{p}/2}}$}
\newcommand{\tdhalf}{$t_{d_{L_{p}/2}}$}

\begin{document}
\let\WriteBookmarks\relax
\def\floatpagepagefraction{1}
\def\textpagefraction{.001}

\shorttitle{Magnetar-driven Light Curves of SESNe}

\shortauthors{Kumar, Amit 2024}

\title [mode = title]{Insights from Modeling Magnetar-driven Light Curves of Stripped-envelope Supernovae}                      

\author[1, 2]{Amit Kumar}[type=author,
                        auid=000,bioid=1,
                        orcid=0000-0002-4870-9436]

\cormark[1]

\ead{amitkundu515@gmail.com; amit.kumar@rhul.ac.uk}

\affiliation[1]{organization={Department of Physics, Royal Holloway - University of London},
    addressline={Egham Hill}, 
    city={Egham},
    postcode={TW20 0EX}, 
    country={UK}}
    
\affiliation[2]{organization={Department of Physics, University of Warwick},
    addressline={Gibbet Hill Road}, 
    city={Coventry},
    postcode={CV4 7AL}, 
    country={UK}}

\cortext[cor1]{Corresponding author}

\begin{abstract}
This work presents the semi-analytical light curve modelling results of 11 stripped-envelope SNe (SESNe), where millisecond magnetars potentially drive their light curves. The light-curve modelling is performed utilizing the $\chi^2$-minimisation code {\tt MINIM} considering millisecond magnetar as a central engine powering source. The magnetar model well regenerates the bolometric light curves of all the SESNe in the sample and constrains numerous physical parameters, including magnetar's initial spin period (\spin) and magnetic field ($B$), explosion energy of supernova (\eexp), progenitor radius (\rp), etc. Within the sample, the superluminous SNe 2010kd and 2020ank exhibit the lowest $B$ and \spin{} values, while the relativistic Ic broad-line SN 2012ap shows the highest values for both parameters. The explosion energy for all SESNe in the sample (except SN 2019cad), exceeding $\gtrsim$2 $\times$ 10$^{51}$ erg, indicates there is a possibility of a jittering jet explosion mechanism driving these events. Additionally, a correlation analysis identifies linear dependencies among parameters derived from light curve analysis, revealing positive correlations between rise and decay times, \spin{} and $B$, \spin{} and \rp, and \eexp{} and \rp, as well as strong anti-correlations of \spin{} and $B$ with the peak luminosity. Principal Component Analysis is also applied to key parameters to reduce dimensionality, allowing a clearer visualization of SESNe distribution in a lower-dimensional space. This approach highlights the diversity in SESNe characteristics, underscoring unique physical properties and behaviour across different events in the sample. This study motivates further study on a more extended sample of SESNe to look for millisecond magnetars as their powering source.
\end{abstract}


\maketitle

\section{Introduction}\label{sec:intro}

Stripped-envelope supernovae (SESNe) originate from the core-collapse of massive stars ($>$8 M$_{\odot}$; \citealt{Heger2003, Smartt2009}) that have stripped off their outer H/He-envelopes before the explosion, e.g., massive Wolf–Rayet stars ($\gtrsim$\,20\,M$_\odot$; \citealt{Woosley1995, Georgy2009, Yoon2015}) or low-mass stars in binary systems ($\gtrsim$\,11\,M$_\odot$; \citealt{Yoon2010, Smith2011,Smith2014, Solar2024}). Several potential mechanisms exist by which SESNe could stripped off their outer H/He-envelopes, such as steady line-driven winds \citep{Conti1975, Puls2008, Pauldrach2012, Bjorklund2023, Michel2023}, eruptive episodes of mass-loss \citep{Smith2006, Beasor2020, Bonanos2024} and/or mass-transfer to a binary companion through Roche Lobe overflow \citep{Nomoto1984, Podsiadlowski2004, Yoon2010, Lyman2016, Gilkis2019, Drout2023}.

SESNe can be distinguished from Type II and Type Ia SNe respectively through the absence of Balmer and Si{\sc~ii} features in their early spectra (\citealt{Filippenko1997, Gal-Yam2017, Prentice2017}). Further classification of SESNe leads to subclasses, such as Type Ib and Ic, based on the presence and absence of He-features in their spectra, respectively. Type Ic SNe can further be subdivided into superluminous SNe (SLSNe-I), Ic broad-line (Ic-BL), and gamma-ray bursts (GRBs) associated SNe (GRB-SNe); see \citet{Modjaz2019}. SLSNe-I, distinguished by their 10-100 times higher luminosity compared to classical SNe, and W-shaped O{\sc~ii} features in the bluer part of their near-peak optical spectra \citep{Quimby2011, Inserra2019, Nicholl2021, Gomez2024, Moriya2024}. SNe Ic-BL are characterised by broad absorption spectral features, indicating high ejecta expansion velocity (\vexp{} $\gtrsim20,000~km~s^{-1}$; \citealt{Galama1998b, Modjaz2016, Taddia2019}). Some Ic-BL SNe are also found to be associated with GRBs and are thus termed GRB-SNe \citep{Woosley2006, Cano2017, Kumar2024,Kumar2024arXiv241113242K}. There also exists a unique case of an SLSN associated with an ultra-long (UL) GRB 111209A, SN 2011kl \citep{Greiner2015, Kann2019}, opening new avenues for exploring connections between SLSNe-I and GRBs. Additionally, within Type Ic subclasses, there are other peculiar events, such as relativistic Ic-BL SNe with highly relativistic ejecta than those of Ib/c SNe and comparable to sub-energetic GRBs (e.g., SNe 2009bb; \citealt{Levesque2010, Pignata2011}, 2012ap; \citealt{Liu2015, Milisavljevic2015} and 2023adta; \citealt{Siebert2024}), see \cite{Margutti2014}. 

Type Ic-BL SNe are among the most energetic SNe, with explosion energies (\eexp) usually reaching $\gtrsim$10$^{52}$ erg \citep{Maeda2003, Janka2016, Prentice2016, Cano2017}, higher than those of other subtypes of SESNe, those exhibit explosion energies in the range of $\approx$10$^{50-52}$ erg \citep{Utrobin2015, Taddia2018, Gal-Yam2019, Burrows2021}. Studies have shown that the explosion energy of SNe can be helpful in probing their explosion mechanism (\citealt{Shishkin2023}, see also \citealt{Soker2024b} for a recent review). The two popular theoretical mechanisms that explain the explosion of SESNe are the delayed neutrino explosion mechanism \citep{Bethe1985, Burrows2020, Bollig2021, Burrows2021, Boccioli2024, Yamada2024} and the jittering jet explosion mechanism (JJEM; \citealt{Soker2010, Papish2011, Gilkis2015, Antoni2022, Soker2022a,Soker2023a, Bear2025}). Whereas, the delayed neutrino explosion mechanism can not explain the SNe with \eexp{} $\gtrsim$2 $\times$ 10$^{51}$ erg \citep{Fryer2012, Sukhbold2016, Gogilashvili2021}. Hence, the exceptionally high explosion energy of most of the Type Ic-BL SNe suggests that JJEM may be the more favourable explosion mechanism for most Ic-BL SNe and in general most core-collapse SNe \citep{Soker2022b}.

Furthermore, within the SESNe class, several intermediate events lie between Type I and Type II SNe, including IIb, Ibn and Icn. Type IIb events initially show H-lines that diminish over time, eventually revealing prominent He-features, such as SN 1993J \citep{Woosley1994}. Type Ibn SNe share similarities with Type Ib (lack of H{\sc~i} and Si{\sc~ii} features) but exhibit additional characteristics of He-rich circumstellar material (CSM)-ejecta interaction, e.g., SN 2006jc \citep{Pastorello2007}. Similarly, Type Icn SNe resemble typical Type Ic events but manifest additional interaction signatures from carbon- or oxygen-rich CSM-ejecta interaction, as observed in SNe 2019hgp \citep{Gal-Yam2022} and 2021csp \citep{Perley2022}. Whereas, the recent discovery of transitional events like SN 2022crv, which exhibits characteristics between Types Ib and IIb, further complicates classification within these boundaries \citep{Dong2023}.

The diversity in the observed properties of different types of SNe and SNe within the same class is crucial to understanding how the lives of stars end differently. In addition to distinct progenitor and environment systems, diversity in the nature of the powering source and the different characteristics of underlying powering sources could play a vital role in governing the distinct properties of SNe. Classical Type Ib/c SNe are thought to be mainly governed by the radio-active decay chain $^{56}$Ni $\rightarrow$ $^{56}$Co $\rightarrow$ $^{56}$Fe \citep[RD model,][]{Arnett1982, Barbon1984}. Some of the SESNe also have circumstellar material (CSM)-ejecta interaction signatures in their early observations, hence claimed to have contributions from the energy deposition from the CSM interaction (CSMI model), e.g., LSQ13ddu \citep{Clark2020}, SN 2019tsf and SN 2019oys \citep{Sollerman2020}. On the other hand, some of Ib/c SNe with peculiar light curves are claimed to be powered by newly formed millisecond magnetar as a central-engine-based power source (MAG model, \citealt{Kasen2010, Woosley2010}) or a combined effect involving both millisecond magnetar and $^{56}$Ni-decay, e.g.,  SNe 1997ef \citep{Wang2016}, 2005bf \citep{Maeda2007}, 2007ru \citep{Wang2016}, PTF11mnb \citep{Taddia2018}, and 2019cad \citep{Gutierrez2021}, see also recent study by \cite{Hu2023}. In addition, spin-down millisecond magnetar is one of the most favourable models to explain the unique observational properties of most SLSNe-I \citep{Inserra2013, Nicholl2017, Dessart2019, Gottlieb2024} and GRB-SNe \citep{Kumar2024}. Additionally, recently \cite{Osmar2024} have investigated central-engine-based power sources to explain the light curves of SESNe.

Conversely, it's challenging to explain the doubled peak light curves and post-peak undulations in the light curves of SESNe (mainly SLSNe-I) by considering magnetar as a single powering source \citep{Nicholl2015a,Nicholl2016b}. However, some studies suggested that introducing a shock breakout driven by the magnetar can explain the initial peak \citep{Kasen2016, Liu2021}, and variable thermal energy injection in the ejecta from a centrally-located magnetar can explain the post-peak undulations in the light curves of SESNe \citep{Moriya2022}, see also \cite{Zhu2024}. Furthermore, the inclusion of jets as a powering source in addition to the magnetar model is found effective in explaining the light curves (along with their bumps) of most core-collapse SNe, including luminous SNe and SLSNe-I (\citealt{Soker2017, Soker2022d, Gottlieb2024}, see \citealt{Soker2022c} for a review). Additionally, the possibility of the formation of jets in fast-rotating progenitors and observed GRBs originated from such progenitors also raise the possibility of contributions of jets in powering the GRB-SNe \citep{Corsi2021, Soker2022c}. However, we would like to caution that the magnetar model with the {\tt MINIM} code \citep{Chatzopoulos2013} used in the present study does not include the above-mentioned features and is unable to fit the pre-peak bumps and the post-peak undulations.

The present study investigates a sample of 11 SESNe with light curves potentially driven by spin-down millisecond magnetars, employing a unified approach to constrain the physical parameters. Additionally, it examines inter-relationships among the derived parameters and also investigates the distribution of SESNe in multi-dimensional parameter space. The paper is organized as follows: Section~\ref{sec:sample} provides an overview of the SESNe sample used in this study. Section~\ref{sec:LC_modelling} details the light curve modelling procedures, along with the derived fitting parameters and results. Section~\ref{sec:LC_parms_inves} presents the linear correlations among SESNe parameters obtained from the light curve analysis and investigates their distribution in reduced dimensionality through Principal Component Analysis. Finally, Section~\ref{sec:summary} summarizes the key findings.

\section{Sample}
\label{sec:sample}

This work investigates millisecond magnetars as powering sources in shaping the light curves of various subtypes of SESNe through light curve modelling. The sample comprises a total of 11 SESNe, which includes 1 Ib (SN 2012au), 1 Ibc (SN 2005bf), 4 Ic (SNe 1997ef, 2007ru, PTF11mnb and 2019cad), 1 relativistic Ic-BL (SN 2012ap), 1 GRB-SN (SN 1998bw), 1 GRB-SLSN (SN 2011kl) and 2 SLSNe-I (SNe 2010kd and 2020ank). This section discusses the motivation behind choosing these specific events and shares some details about these cases.

\textbf{\textit {SLSNe-I:}} The majority of SLSNe-I are explained considering the magnetars as their primary power sources \citep{Nicholl2017, Yu2017}. From the SLSNe-I category, SN 2010kd (slow-evolving) and SN 2020ank (fast-evolving) are added to the sample because the bolometric light curves of SNe 2010kd and 2020ank are regenerated using the magnetar model under the {\tt MINIM} code and employing the same assumptions (see \citealt{Kumar2020} and \citealt{Kumar2021}, respectively), which are being used in the present work. Hence, these candidates are the most suitable for the sample from the SLSNe-I category, ensuring coherence and consistency. Furthermore, for SNe 2010kd and 2020ank, in addition to the indication of spin-down millisecond magnetars as power sources from the {\tt MINIM}-based light-curve modelling, the high UV-peak brightness of both events and the flatter expansion velocity in SN 2010kd suggest a central-engine-based powering mechanism for these objects \citep{Kumar2020, Kumar2021}.

\textbf{\textit { GRB-SLSN:}} To date, SN 2011kl is the only SLSN associated with a ULGRB 111209A, and it is also an intermediate event with a brightness higher than all classical GRB-SNe and lower than SLSNe-I \citep{Greiner2015, Kann2019}. Previous studies recommended MAG \citep{Greiner2015, Gompertz2017, Kumar2022,Kumar2024} or MAG+RD \citep{Bersten2016, Wang2017a} models to explain the unique properties of SN 2011kl. Whereas, \citet{Gao2016} indicated mass-accreting black as a central engine powering source for SN 2011kl, whereas \citet{Lin2020} proposed the formation of a supermassive magnetar which later collapsed into a black hole. \citet{Kumar2022,Kumar2024} also reproduced the light-curve of SN 2011kl using the MAG model and employing the {\tt MINIM} code. The signatures of millisecond magnetar as a primary powering source of SN 2011kl, its association with an ULGRB, and intermediate luminosity between SLSNe-I and GRB-SNe make it a promising candidate for the sample.

\textbf{\textit { GRB-SN:}} SN~1998bw is the first and most nearby (z = 0.00866) Ic-BL SN found associated with GRB 980425 thus far \citep{Galama1998b, Iwamoto1998, Wang1998, Patat2001}. It was termed a hypernova because its kinetic energy was very high, nearly tenfold higher than that of conventional Type Ic SNe. Additionally, the accompanying GRB 980425 displayed a lower isotropic $\gamma-$ray luminosity in comparison to other cosmological LGRBs \citep{Kulkarni1998}. The late-time photometric evolution of SN 1998bw exhibited a faster decline than the decay rate of $^{56}$Co $\rightarrow$ $^{56}$Fe, whereas the late-time spectra lack ejecta-wind interaction signatures \citep{Patat2001}, those disfavour RD and CSMI as dominant power sources. On the other hand, \citet{Wang2017} and \citet{Kumar2024} proposed MAG+RD and MAG, respectively, as the most suitable models to explain the light curve of SN 1998bw. Given its proximity and extensive observations, SN 1998bw emerges as a compelling addition to the sample from the GRB-SNe family.

\textbf{\textit { Relativistic Ic-BL SN:}} SN 2012ap exhibits mildly relativistic ejecta, with near-peak expansion velocity comparable to SN 1998bw. Although there is no observational signature of association of a GRB with SN 2012ap, it is an interesting object because of its highly relativistic behaviour similar to those of GRB-SNe \citep{Margutti2014, Chakraborti2015, Milisavljevic2015, Liu2015}. Notably, its radio observations designate SN 2012ap as the least energetic explosion driven by a central engine \citep{Margutti2014, Chakraborti2015}. Moreover, \citet{Kumar2024} suggested a millisecond magnetar as the primary power source for SN 2012ap, supported by light curve modelling using the {\tt MINIM} code. The above characteristics make SN 2012ap a good candidate to add to the sample.

\textbf{\textit { Type Ic SNe:}} Within the Type Ic category, the sample includes SNe 1997ef, 2007ru, PTF11mnb and 2019cad, all suggested to be powered by millisecond magnetars in different literature.

{\it SN 1997ef:} shares high kinetic energy and a broad peak light curve similar to SN 1998bw but distinct from classical Type Ic SNe \citep{Iwamoto1998, Mazzali2000,Mazzali2004}. Based on light curve modelling, \citet{Wang2016} suggested the MAG+RD as the most suitable model for this event.

{\it SN 2007ru:} exhibits high ejecta expansion velocity and high peak luminosity comparable to SN 1998bw, placing it among the most luminous Type Ic SNe \citep{Sahu2009}. Similar to SN 1997ef, \citet{Wang2016} proposed the MAG+RD model as the optimal choice to trace the bolometric light curve of SN 2007ru.

{\it PTF11mnb:} displays a double-peak light curve with a fainter first peak followed by a more prominent secondary peak \citep{Taddia2018}, similar to SN 2005bf \citep{Folatelli2006}. In their analysis, \citet{Taddia2018} characterised the first peak using the RD model, while the main/secondary peak was defined using the MAG model. The light curve was also interpreted with a model accounting for a double-peaked $^{56}$Ni distribution \citep{Taddia2018}.

\begin{figure}
\centering
\includegraphics[angle=0,scale=0.8]{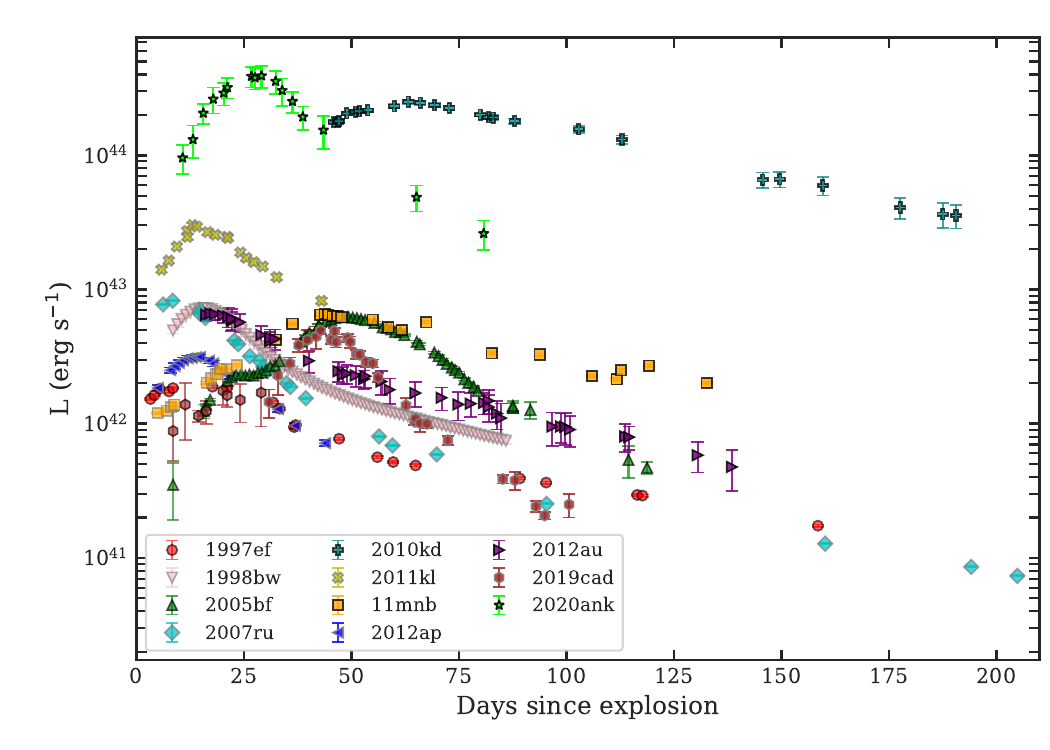}
\caption{Bolometric light curves of the full sample of SESNe used in the present work are shown. Data courtesy: SN 2010kd \citep{Kumar2020}, SN 2020ank \citep{Kumar2021}, SN 2011kl, SN 1998bw, SN 2012ap \citep[][and references threin]{Cano2017}, SN 1997ef \citep{Iwamoto2000, Mazzali2000,Mazzali2004}, SN 2019cad\citep{Gutierrez2021}, SN 2005bf \citep{Anupama2005, Tominaga2005, Folatelli2006}, SN 2007ru \citep{Sahu2009}, PTF11mnb \citep{Taddia2018}, SN 2012au \citep{Pandey2021}.}
\label{fig:bolo_LC}
\end{figure}

\begin{figure*}
\centering
\includegraphics[angle=0, scale=0.46]{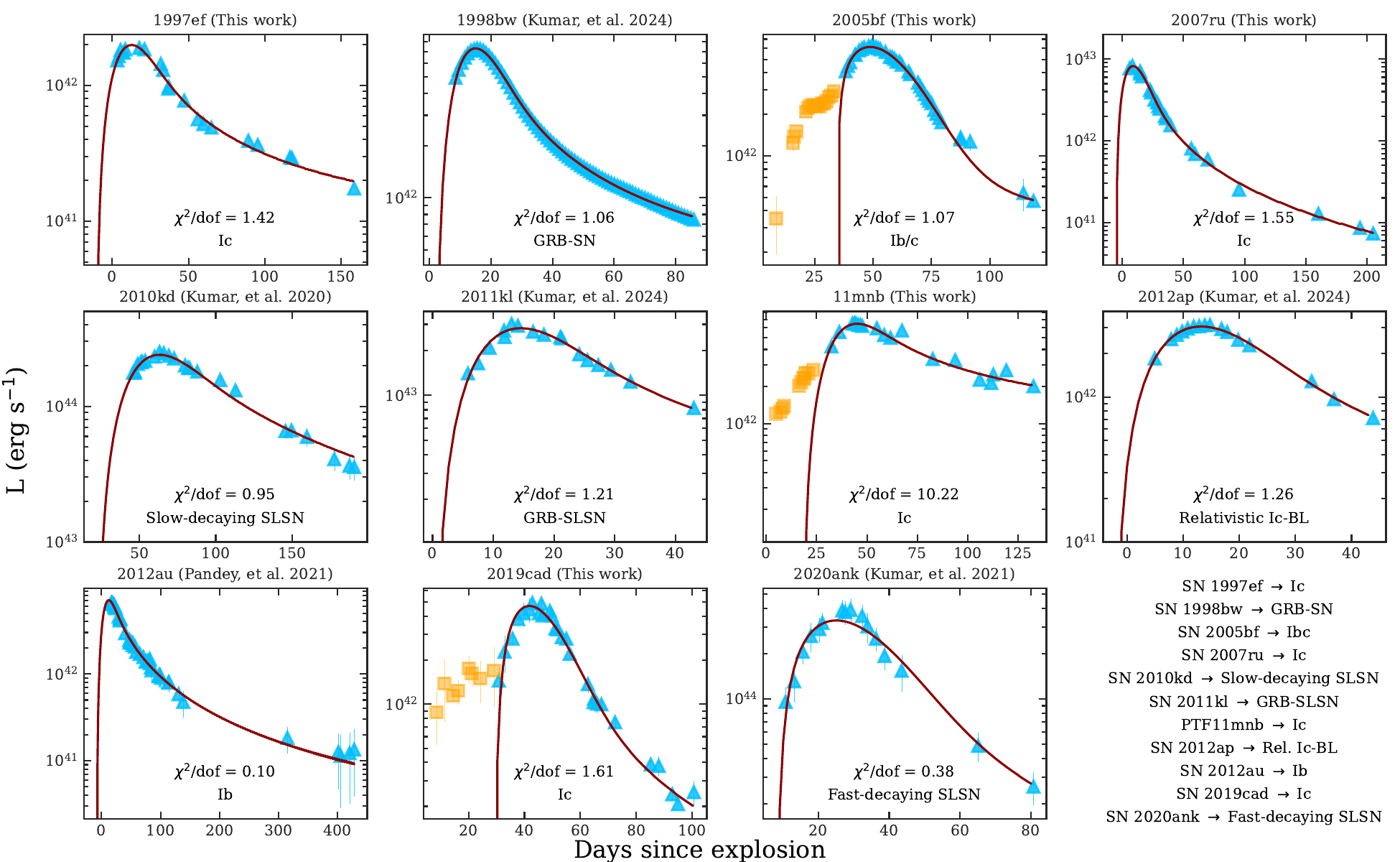}
\caption{Bolometric light curve fittings of 11 SESNe performed utilising the MAG model within the \texttt{MINIM} code \citep{Chatzopoulos2013} are shown. The light curves of all SESNe within the sample are fitted with low $\chi^{2}/\mathrm{dof}$ values (except PTF11mnb, due to highly scattered data). Pre-peak bumps (in orange) in the case of SNe 2005bf, PTF11mnb and 2019cad are masked and not included in the fitting. See Table~\ref{tab:mag_parameters} for estimated fitting parameters.}
\label{fig:light_curve_fitting}
\end{figure*}

{\it SN 2019cad:} shows a dual-peak light curve similar to PTF11mnb \citep{Gutierrez2021}. Therefore, akin to PTF11mnb, the bolometric light curve of SN 2019cad is explained using a double-peaked $^{56}$Ni distribution and a hybrid model which involves tracing the first peak using the RD model and the main peak using the MAG model \citep{Gutierrez2021}.

\textbf{\textit { Type Ibc SN:}} SN 2005bf exhibits a dual-peak light curve and is claimed a Type Ibc SN with a possible asymmetric ejected envelope \citep{Anupama2005, Tominaga2005, Folatelli2006, Maund2007, Parrent2007}. Its early features indicate it as a type Ic SN; however, the late phase spectral evolution seems similar to Type Ib SNe. \cite{Maeda2007} suggested that the shock heating could power the low-luminous/first peak, whereas a millisecond magnetar may be the primary power source for the secondary/main peak.

\textbf{\textit { Type Ib SN:}} SN 2012au is a very peculiar Type Ib SN, an intermediate event between Ib/c and Ic-BL SNe \citep{Pandey2021}. No CSM interaction signatures were found in the very late-time spectrum of SN 2012au, and a pulsar wind nebula remnant was suggested for this event \citep{Milisavljevic2018}. In addition, this is the only Type Ib SN that is claimed to be driven by a millisecond magnetar based on light-curve \citep{Pandey2021} and spectral modelling \citep{Omand2023}. These properties prove SN 2012au is an excellent event for the sample.

This study uses this diverse sample of SESNe to investigate the magnetar as a primary central-engine powering source among various subtypes of SESNe. Bolometric light curves were generated only for certain SESNe in the sample, as needed, using the method detailed in \cite{Kumar2024}. The bolometric light curves of all the 11 SESNe mentioned above are shown in Figure~\ref{fig:bolo_LC}. Data courtesy: SN 2010kd \citep{Kumar2020}, SN 2020ank \citep{Kumar2021}, SN 2011kl, SN 1998bw, SN 2012ap \citep[][and references threin]{Cano2017}, SN 1997ef \citep{Iwamoto2000, Mazzali2000,Mazzali2004}, SN 2019cad\citep{Gutierrez2021}, SN 2005bf \citep{Anupama2005, Tominaga2005, Folatelli2006}, SN 2007ru \citep{Sahu2009}, PTF11mnb \citep{Taddia2018}, SN 2012au \citep{Pandey2021}. The bolometric light curves for the events within this sample cover a wide range of peak luminosity from lowest peak luminosity $\sim2 \times 10^{42}$ erg s$^{-1}$ (for SN 1997ef) to highest peak luminosity $\sim4 \times 10^{44}$ erg s$^{-1}$ (for SN 2020ank).

\section{Light curve modelling}
\label{sec:LC_modelling}

This work focuses on investigating millisecond magnetars as the primary power source governing the light curves of SESNe and constrains the crucial parameters. To achieve this, {\tt MINIM} code is used, which is a promising tool to perform the semi-analytical light curve modelling of SESNe \citep{Chatzopoulos2013}, see also \citep{Wheeler2017, Kumar2020,Kumar2021, Pandey2021, Kumar2022,Kumar2024}. {\tt MINIM} is a robust $\chi^2$-minimization tool implemented in {\tt C++} \citep{Chatzopoulos2013}. A comprehensive description of {\tt MINIM}, including its operational methodology, is detailed in \citet{Chatzopoulos2013}.

To probe the underlying power source, {\tt MINIM} code facilitates the performance of light curve modelling of SNe using numerous models, e.g., RD, CSMI, MAG, and hybrid (CSM+RD) models. However, the present work explores the spin-down millisecond magnetar as a central-engine-based power source for a sample of SESNe; therefore, this study focuses only on the MAG model. For a comprehensive understanding of the MAG model and the underlying assumptions applied in the current study, see \citet{Chatzopoulos2012, Chatzopoulos2013}. 

Under the MAG model, there are six fitting parameters: $t_\textrm{i}$ (the initial epoch of explosion), magnetar rotational energy (\ep), diffusion time scale (\td), spin-down timescale (\tp), progenitor radius (\rp), and \vexp. Whereas, other crucial parameters such as initial spin period (\spin), magnetic field ($B$), ejecta mass (\mej) and explosion energy of the SN (\eexp) can be estimated utilising the above parameters obtained through light curve fitting (e.g., \ep, \tp, \td{} and \vexp). As outlined in \citep{Chatzopoulos2012,Chatzopoulos2013}, $P_{\rm i} = (\frac{2 \times 10^{50}\,{\rm erg}\,{\rm s}^{-1}} {E_{\rm p}})^{0.5} \times 10 $ ms and $B = (\frac{1.3 \times P_{10}^{2}}{t_{\rm p,yr}})^{0.5} \times 10^{14}$ G. Furthermore, the values of \mej{} can be computed using equation 1 of \citet{Wheeler2015}, $M_{\rm ej} \sim \frac{1}{2}\frac{\beta c}{\kappa} V_{\rm exp} t_{\rm d}^2$, adopting $\beta$ = 13.8 and $\kappa$ = 0.1 cm$^2$ g$^{-1}$, considering half fully-ionized gas, as recommended by \citet{Inserra2013, Wheeler2015, Wang2017}. In addition, \eexp{} can be estimated using equation 3 of \cite{Wheeler2015}, $E_{\rm SN} \approx \frac{3}{10} M_{\rm ej} V_{\rm exp}^2 $, considering kinetic energy of the SN approximately equals to the explosion energy.

The MAG model within the {\tt MINIM} code well-explained the light curves of all the SESNe presented here, with $\chi^2$/dof values close to one, except for PTF11mnb ($\chi^2$/dof = 10.22), potentially due to large scatter in its data. For five SESNe (1997ef, 2005bf, 2007ru, PTF11mnb, and SN 2019cad), the light curve modelling using the {\tt MINIM} code is conducted independently within this work. In the case of, SNe 2005bf, PTF11mnb and 2019cad, strategically masked the initial/low-luminosity peak (depicted in orange colour in Figure~\ref{fig:light_curve_fitting}) and focused on the light curve modelling solely on the secondary/main peak, as suggested by \citet{Maeda2007}, \citet{Taddia2018} and \citet{Gutierrez2021}, respectively. On the other hand, for the rest of six SNe 2010kd \citep{Kumar2020}, 2020ank \citep{Kumar2021}, 2011kl, 1998bw, 2012ap \citep{Kumar2024}, and 2012au \citep{Pandey2021}, the light curve modelling results are directly adopted from existing literature, where the {\tt MINIM} code was employed to model the light curves, ensuring methodological consistency in this analysis. Nevertheless, only the \mej{} values are recalculated for these cases by incorporating the provided equation and assuming a consistent $\kappa$ value. It is important to acknowledge that a degree of parameter degeneracy could introduce potential bias in the inferred parameters, as extensively discussed by \citet{Chatzopoulos2013}; nevertheless, careful consideration is affirmed in interpreting the results.

The results of the light curve fitting are shown in Figure~\ref{fig:light_curve_fitting}, and the corresponding fitting parameters are listed in Table~\ref{tab:mag_parameters}. Table~\ref{tab:mag_parameters} includes only the parameters directly obtained from the light curve modelling (\ep, \tp, \td, \rp, and \vexp), whereas $t_\textrm{i}$ is not included due to large uncertainty associated because of masking initial faint peak in some cases. While parameters derived from these fitting parameters, such as \spin, $B$, \mej, and \eexp{} are tabulated in Table~\ref{tab:ligh_curve_parms}. In addition, Table~\ref{tab:ligh_curve_parms} also listed peak luminosity (\lp), rise time from half-peak to peak luminosity during the pre-peak phase (\trhalf), and decay time from peak to half-peak luminosity in the post-peak phase (\tdhalf) for SESNe in the sample, that capture the light curve behaviour of SNe around the peak. These parameters were determined via fitting using the Gaussian Process (GP) regression \citep{Rasmussen2005, Bishop2006} with a Radial Basis Function (RBF) kernel, offering a more precise fit than spline fitting and replicating the bolometric light curves while naturally providing phase-based uncertainty estimates \citep{Inserra2018a, Kumar2024arXiv241113242K}. The Python packages {\tt sklearn} and {\tt scipy} were employed for the analysis. For cases lacking pre-peak data near half-peak phases (except for SNe 2011kl, 2019cad, and 2020ank), \trhalf{} was inferred from the magnetar fitting curve.

As expected, SLSNe 2010kd and 2020ank exhibit $\approx$10 - 100 times higher $L_{p}$ in comparison to the other SESNe studied here. In addition, slow-evolving SLSN 2010kd shows a longer rise time of approximately 23.7 d, while SNe Ib/c, such as 2005bf and 2007ru, exhibit shorter rise times, around 8-12 d, which suggests a difference in energy injection timescales among these objects. Within the sample used here, SLSNe and Ibc SNe show comparable values of \vexp{} ($\sim$12,000 km s$^{-1}$), while Ic-BL SNe 1998bw and 2012ap, and GRB-SLSN 2011kl, exhibit comparatively higher \vexp{} ($\gtrsim$20,000 km s$^{-1}$). The \rp{} varies widely, with SLSNe 2010kd and 2020ank having smaller radii around $10^{13}$ cm, while the progenitor radii of SNe Ib/c, such as PTF11mnb, extend to around $15 \times 10^{13}$ cm. Among SNe Ib/c (2005bf, 2007ru, PTF11mnb, 2012au, and 2019cad; excluding 1997ef), there is consistency in \spin{} values around 20 ms, whereas SLSNe 2010kd and 2020ank share the lowest \spin{} values around 2.3 ms. On the other hand, SNe Ib/c events show higher $B$ values, with SNe 2005bf and 2012ap exhibiting strengths around $23 \times 10^{14}$ G and $34 \times 10^{14}$ G, respectively, while SLSNe 2010kd and 2020ank display comparatively lower values of $B$ around $0.8 \times 10^{14}$ G and $2.9 \times 10^{14}$ G, respectively.

Furthermore, within the sample of SESNe used in the present study, the \eexp{} is higher in GRB-SNe (1998bw and 2011kl) and Rel. Ic-BL SN (2012ap), with \eexp{} $>10^{52}$ erg, in contrast to lower values observed in other SESNe, with SN 2019cad with the lowest \eexp{} of $\sim$1.57 $\pm$ 0.33 $\times$ $10^{51}$ erg. However, a detailed study on SN 2019cad by \cite{Gutierrez2021} suggested \eexp{} of 3.5 $\times$ $10^{51}$ erg for this event. This discrepancy could be because of the different models used to reproduce the light curve of SN 2019cad. As discussed above, among the two popular theoretical mechanisms that explain the explosion of SESNe - delayed neutrino explosion and JJEM - the former can not explain the SNe with \eexp{} $\gtrsim$2 $\times$ 10$^{51}$ erg \citep{Fryer2012, Sukhbold2016, Gogilashvili2021}. All the SESNe in the sample used here show \eexp{} $>$2 $\times$ 10$^{51}$ erg, indicating JJEM as their possible explosion mechanism and indicating that these SESNe could be driven by jets (see a review by \citealt{Soker2022c} for details jet-driven core-collapse SNe).

\begin{table*}
\begin{center}
\begin{threeparttable}
\scriptsize
\caption{Optimal parameters along with their associated errors provided within the parentheses for 11 SESNe in the sample obtained through the light curve modelling under the MAG model and employing the {\tt MINIM} code \citep{Chatzopoulos2013}.}
\addtolength{\tabcolsep}{0pt}
\label{tab:mag_parameters}
\begin{tabular}{m{3em} m{5em} m{6em} m{6em} m{6em} m{6em} m{6em} m{2em} m{10em}}
\hline
object & Type & $E_\textrm{p}^a$ & $t_\textrm{d}^b$ & $t_\textrm{p}^c$ & $R_\textrm{p}^d$ & $V_\textrm{exp}^e$ & $\frac{\chi^2}{dof}$ & Source \\
 & & ($10^{49}$~erg) & (d) & (d) & (10$^{13}$~cm) & (10$^{3}$~km/s) & \\ 
\hline
1997ef & Ic  & 1.29 (0.01) & 21.53 (0.21) & 18.70 (0.36) & 12.42 (1.18) & 11.98 (2.38) & 1.42 & This work \\
1998bw & GRB-SN & 2.67 (0.01) & 12.35 (0.07) & 9.49 (0.09) & 19.26 (0.57) & 30.81 (1.43) & 1.06 & \citet{Kumar2024} \\
2005bf & Ibc  & 5.89 (0.11) & 26.96 (0.15) & 2.98 (0.11) & 12.25 (0.64) & 10.87 (0.67) & 1.07 & This work \\
2007ru & Ic & 4.72 (0.01) & 15.86 (0.06) & 6.01 (0.03) & 0.28 (0.03) & 14.02 (0.38) & 1.55 & This work \\
2010kd & SLSN & 337 (6) & 35.0 (0.88) & 46.85 (2.08) & 1.0 (1.25) & 11.2 (0.98) & 0.95 & \citet{Kumar2020} \\
2011kl & GRB-SLSN & 11.87 (0.71) & 12.68 (0.32) & 12.70 (1.22) & 10.26 (3.56) & 24.46 (2.46) & 1.21 & \citet{Kumar2024} \\
11mnb  & Ic & 4.61 (0.36) & 20.57 (0.41) & 37.57 (1.72) & 14.81 (0.99) & 11.41 (0.35) & 10.22 & This work \\
2012ap & Rel. Ic-BL & 1.20 (0.01) & 16.17 (1.07) & 6.71 (0.21) & 29.63 (4.35) & 21.38 (4.56) & 1.26 & \citet{Kumar2024} \\
2012au & Ib & 6 (1) & 20.83 (2.28) & 24.49 (2.12) & 0.36 (0.13) & 11.66 (0.58) & 0.10 & \citet{Pandey2021} \\
2019cad & Ic & 4.14 (0.11) & 18.22 (0.22) & 2.07 (0.08) & 0.04 (0.27) & 10.10 (0.91) & 1.61 & This work \\
2020ank & SLSN &402 (19) & 25.01 (0.16)  & 2.79 (0.35) & 0.28 (0.22) & 12.27 (0.91) & 0.38 & \citet{Kumar2021} \\
\hline
\end{tabular}
    \begin{tablenotes}[para,flushleft]
        $^a$ $E_\textrm{p}$: initial rotational energy of the magnetar.
        $^b$ $t_\textrm{d}$: diffusion timescale.
        $^c$ $t_\textrm{p}$: spin-down timescale.    
        $^d$ $R_\textrm{p}$: progenitor star's radius.
        $^e$ $V_\textrm{exp}$: expansion velocity.
    \end{tablenotes}
  \end{threeparttable}
  \end{center}
\end{table*}

\begin{table*}
\begin{center}
\begin{threeparttable}
\scriptsize
\caption{Light curve parameters for all 11 SESNe in the sample, along with their associated errors (provided in parentheses), estimated using the parameters listed in Table~\ref{tab:mag_parameters} and Gaussian Process (GP) regression fitting on bolometric light curves.}
\addtolength{\tabcolsep}{0pt}
\label{tab:ligh_curve_parms}
\begin{tabular}{m{3em} m{5em} m{6em} m{6em} m{6em} m{6em} m{6em} m{6em} m{6em}}
\hline
object & Type & \spin{}$^a$ & $B^b$ & \mej{}$^c$ & $E_\textrm{exp}^d$ & $L_\textrm{p}^e$ & $t_{r_{L_{p}/2}}^f$ & $t_{d_{L_{p}/2}}^g$ \\
 & & (ms) & ($10^{14}$~G) & ($M_{\odot}$) & ($10^{51}$~erg) & ($10^{42}$~erg/s) & (d) & (d) \\ 
\hline
1997ef & Ic  & 39.31 (0.08) & 19.83 (0.19) & 4.28 (0.94) & 3.66 (1.66) & 1.93 (0.06) & 13.96 & 21.79 (5.73) \\
1998bw & GRB-SN & 27.38 (0.04) & 19.38 (0.09) & 3.62 (0.21) & 20.50 (2.24) & 7.26 (0.03) & 7.81 & 13.80 (0.60) \\
2005bf & Ibc  & 18.42 (0.18) & 23.29 (0.41) & 6.08 (0.20) & 4.29 (0.55) & 6.23 (0.03) & 11.77 & 21.58 (0.86) \\
2007ru & Ic & 20.58 (0.04) & 18.31 (0.05) & 2.72 (0.09) & 3.19 (0.20) & 8.09 (0.20) & 8.83 & 16.24 (1.16) \\
2010kd & SLSN & 2.44 (0.13) & 0.78 (0.01) & 10.56 (1.46) & 4.27 (0.78) & 246.38 (3.78) & 23.71 & 49.16 (2.70) \\
2011kl & GRB-SLSN & 12.98 (0.39) & 7.95 (0.38) & 3.03 (0.46) & 10.82 (2.72) & 29.22 (1.11) & 9.05 (2.01) & 13.85 (3.07) \\
11mnb  & Ic & 20.82 (0.26) & 7.41 (0.17) & 3.72 (0.26) & 2.89 (0.27) & 6.70 (0.14) & 16.06 & 44.87 (8.64) \\
2012ap & Rel. Ic-BL & 40.77 (0.97) & 34.34 (0.54) & 4.30 (1.49) & 11.73 (6.44) & 3.13 (0.02) & 9.41 & 19.31 (0.59) \\
2012au & Ib & 18.26 (0.01) & 8.05 (0.15) & 3.90 (1.05) & 3.83 (0.92) & 6.74 (0.10) & 11.78 & 23.72 (0.43) \\
2019cad & Ic & 21.97 (0.30) & 33.29 (0.67) & 2.58 (0.29) & 1.57 (0.33) & 4.79 (0.15) & 11.06 (2.08) & 11.41 (1.93) \\
2020ank & SLSN & 2.23 (0.51) & 2.91 (0.07) & 5.91 (0.51) & 3.22 (0.48) & 393.23 (8.29) & 11.94 (0.81) & 11.26 (0.91) \\
\hline
\end{tabular}
\begin{tablenotes}[para,flushleft]
        $^a$ \spin: initial spin period of the magnetar.
        $^b$ $B$: magnetic field strength of magnetar.
        $^c$ \mej: ejected mass.
        $^d$ $E_\textrm{exp}$: explosion energy of the supernova.
        $^e$ $L_\textrm{p}^e$: peak luminosity of bolometric light curve.
        $^f$ $t_{r_{L_{p}/2}}$: rise time  $-$  the time between half peak luminosity to peak luminosity in the pre-peak phase.
        $^g$ $t_{d_{L_{p}/2}}$: decay time  $-$  the time between peak luminosity to half peak luminosity in the post-peak phase.
\end{tablenotes}
\end{threeparttable}
\end{center}
\end{table*}

\section{Investigating Light Curve Parameters}\label{sec:LC_parms_inves}

This section presents the exploratory analysis and dimensionality reduction techniques applied to investigate correlations and dependencies among the parameters of all SESNe in the sample, tabulated in Tables~\ref{tab:mag_parameters} and \ref{tab:ligh_curve_parms}. By examining these sources in multi-dimensional parameter space, the aim is to uncover underlying relationships and gain deeper insights into their behaviour.

\subsection{Correlation Analysis}\label{sec:correlation}

\begin{figure}
\centering
\includegraphics[width=0.8\textwidth]{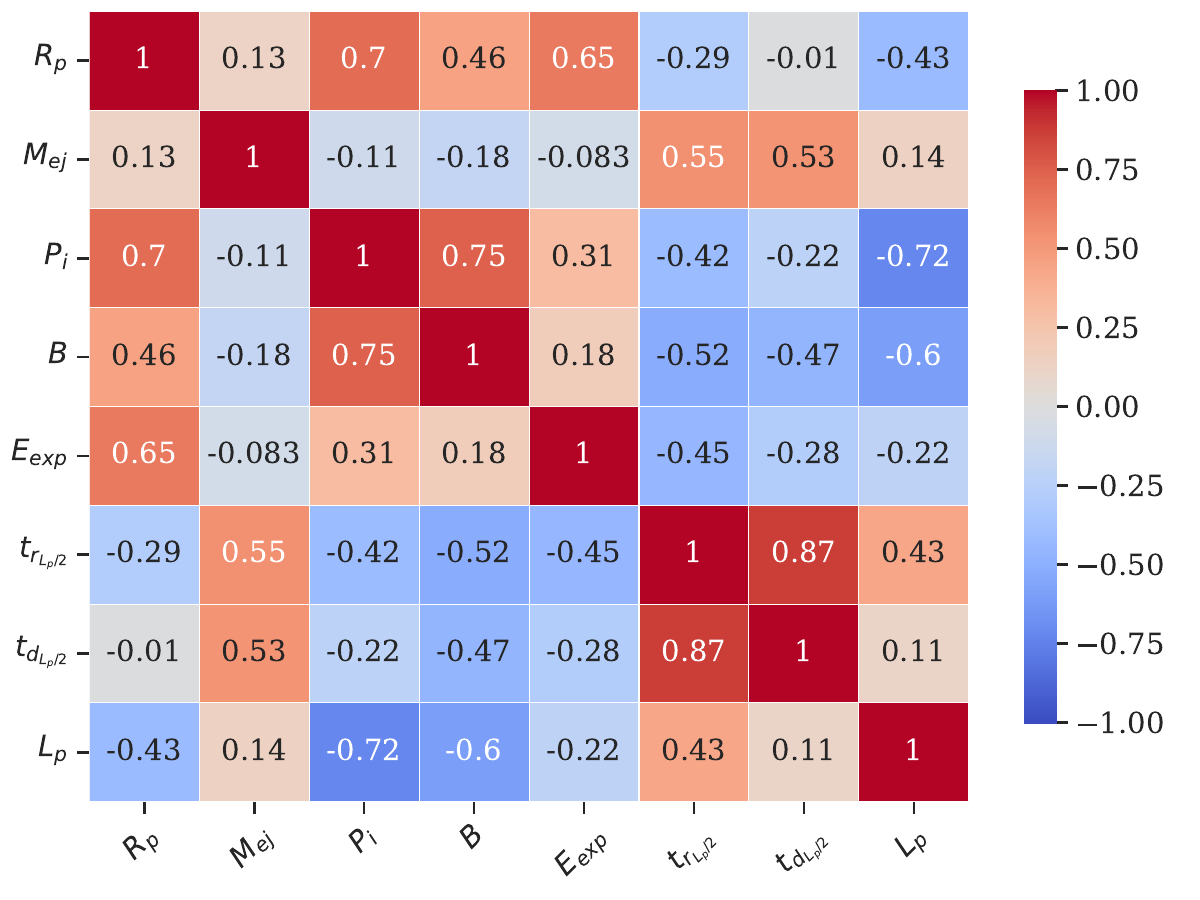}
\caption{Heatmap representation of the correlation matrix among various physical parameters, e.g., $R_p$, \mej, \spin, $B$, \eexp, $t_{r_{L1/2}}$, $t_{d_{L1/2}}$ and $L_p$. The heatmap highlights parameter pairs with strong positive correlations in red, suggesting a linear relationship between those parameters. Whereas, parameters with inverse relationships are shown in blue, providing insight into possible anti-correlation.}
\label{fig:corr_heatmap}
\end{figure}

This section discusses a correlation analysis using Pearson’s correlation coefficient to identify potential linear dependencies among the parameters tabulated in Tables~\ref{tab:mag_parameters} and \ref{tab:ligh_curve_parms}, which includes \rp, \mej, \spin, $B$, \eexp, \trhalf, \tdhalf{} and \lp. In the view of the fact that \mej, \spin, $B$ and \eexp{} are computed directly using the other light curve fitting parameters tabulated in Table~\ref{tab:mag_parameters} (except \rp) as discussed in Section~\ref{sec:LC_modelling}, the parameters \ep, \td, \tp, and \vexp{} are excluded from the correlation analysis. A heatmap of the correlation matrix is presented in Figure ~\ref{fig:corr_heatmap}. 

Parameters $B$ - \rp, \trhalf{} - \mej, \tdhalf{} - \mej and \lp{} - \trhalf{} show the moderate correlations (0.4 to 0.59); and \trhalf{} - \spin, \trhalf{} - $B$, \trhalf{} - \eexp, \tdhalf{} - $B$ and \lp{} - \rp{} present the moderate anti-correlations ($-$0.4 to $-$0.59). Whereas, \spin{} - \rp, \spin{} - $B$, and \eexp{} - \rp{} show strong correlations (0.6 to 0.79); \spin{} - \lp{} and $B$ - \lp{} present strong anti-correlations ($-$0.6 to $-$0.79); and \trhalf{} - \tdhalf{} exhibits very strong correlation (0.87). The details about the parameters that present strong correlation or anti-correlation are as follows.

\begin{figure}
\centering
\includegraphics[angle=0,scale=0.4]{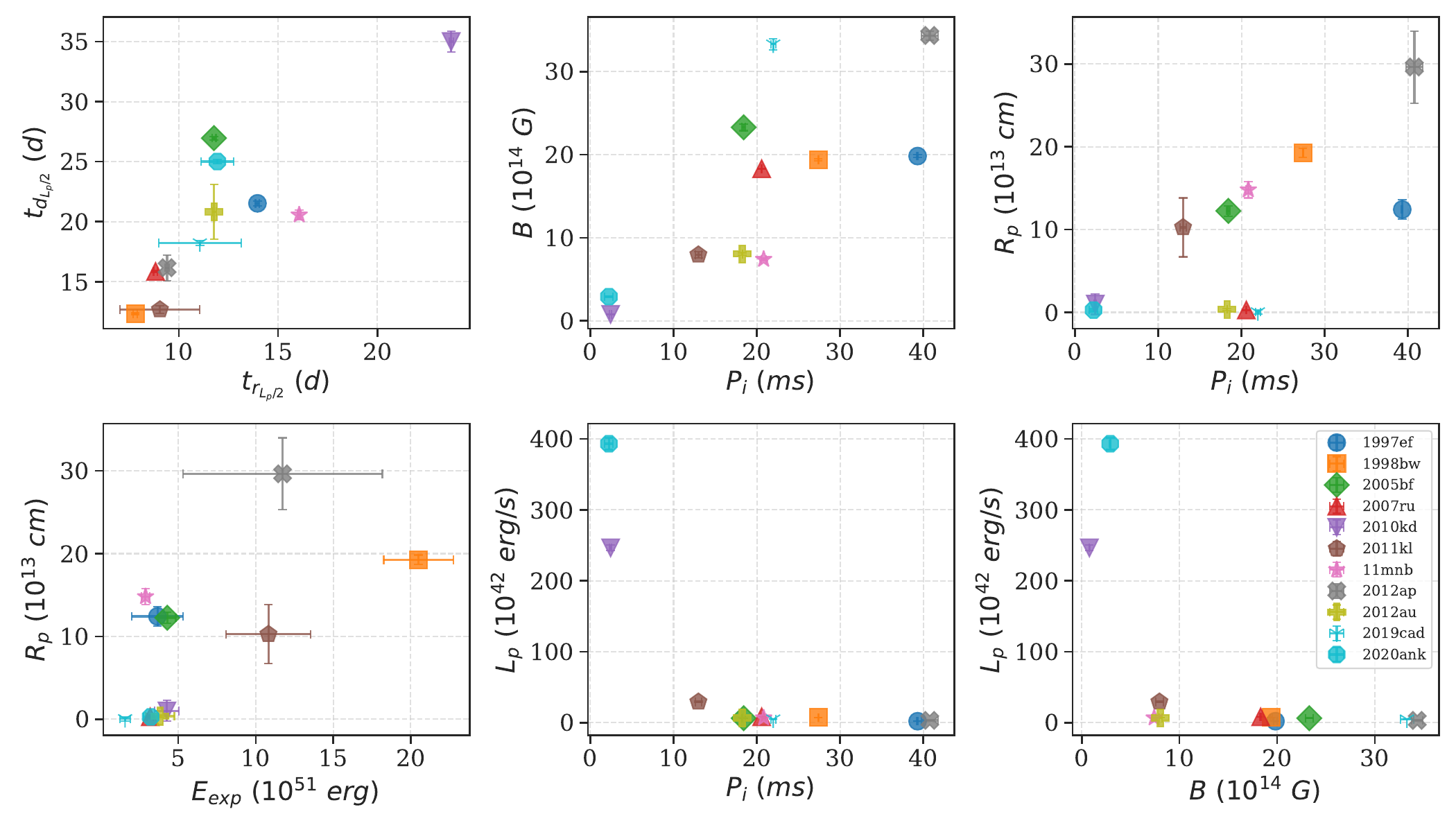}
\caption{Scatter plots for all 11 SESNe in the sample, highlighting parameters with very strong correlations (\trhalf{} vs. \tdhalf), strong correlations (\spin{} vs. $B$, \spin{} vs. \rp, and \eexp{} vs. \rp), and strong anti-correlations (\spin{} vs. \lp{} and $B$ vs. \lp).}
\label{fig:scatter_plot_pairs}
\end{figure}

\begin{itemize}

    \item \trhalf{} $-$ \tdhalf: A very strong positive correlation (0.87) between \trhalf{} and \tdhalf{} indicates that SNe with longer rise rates also tend to have extended decay rates, and vice versa. This relationship stems from the dependency of light curve evolution on the diffusion timescale, which depends on ejecta mass, explosion energy, and opacity \citep{Arnett1982}, see also \citep{Drout2011, Taddia2015, Prentice2016}. Typically, SESNe with longer rise times display slower decay times due to larger ejecta mass, opacity and/or lower explosion energy, which results in extended photon diffusion times. Also, SESNe with higher stripped progenitors and higher explosion energies exhibit sharper rise and decay times due to rapid energy release \citep{Taddia2015, Prentice2016}. The scatter plot of \trhalf{} vs. \tdhalf{} in this study further supports this linear relationship across SESNe subclasses, see the upper-left panel of Figure~\ref{fig:scatter_plot_pairs}. SN 2010kd exhibits the highest rise and decay times (it has the highest diffusion time scale and ejecta mass in the sample) whereas SN 1998bw presents the lowest rise and decay times (it contains the lowest diffusion time and among the low ejecta mass SESNe in the sample), see Tables~\ref{tab:mag_parameters} and \ref{tab:ligh_curve_parms}.\\

    \item \spin{} $-$ $B$: A strong positive correlation between \spin{} and $B$ (0.75) suggests that SESNe within the sample associated with magnetars with higher initial spin periods also exhibits higher magnetic field strengths. The plot presenting \spin{} vs. $B$ in the upper-middle panel of Figure~\ref{fig:scatter_plot_pairs} also shows a linear positive relationship between \spin{} and $B$, though with a significant scatter. Although, as mentioned in Section~\ref{sec:LC_modelling}, $B$ is directly proportional to \spin{} and is inversely proportional to $\sqrt{t_p}$, which could lead to the linear positive relationship observed between $B$ and \spin. Figure 3 of \cite{Kumar2024} also illustrates the \spin{} vs. $B$ relation for different types of transients, including GRB-SNe, SLSNe-I, long GRBs, and short GRBs, showing considerable scatter in \spin{} vs. $B$ even among SNe of the same class (such as in the case of SLSNe-I) and no notable linear relationship can be seen. Thus, the linear relationship between \spin{} and $B$ observed in this study may not be a general trend and could arise from the smaller sample size used in this analysis.
    
    \item \spin{} $-$ \rp: A strong positive correlation (0.7) between \spin{} and \rp{} suggests that SNe originated from progenitors with larger radii and can also form centrally located newborn magnetars with higher initial spin periods (slower rotating). This could be mainly due to rotational energy constraints and angular momentum distribution within the progenitor star. In progenitors with larger radii, angular momentum is distributed over a more significant volume, generally resulting in a slower core rotation rate and a higher initial spin period \citep{Woosley2010, Metzger2015, Muller2019}. The scatter plot of \spin{} vs. \rp{} generated in this work (see the upper-right panel of Figure~\ref{fig:scatter_plot_pairs}) clearly shows a linear relationship between \spin{} and \rp; SESNe with higher \spin{} exhibit higher \rp, however, with four outliers (SNe 1997ef, 2007ru and 2012au, 2019cad).\\

    \item \eexp{} $-$ \rp: A strong positive correlation of 0.65 between \eexp{} and \rp{} suggests that SNe with more extended progenitor tend to have higher explosion energies. The scatter plot between \eexp{} and \rp{} is shown in the lower-left panel of Figure~\ref{fig:scatter_plot_pairs}, where a positive trend is not clearly visible because of the huge scatter. Contrary, \cite{Goldberg2019} demonstrate that the explosion energy of a supernova is inversely proportional to the progenitor radius, whereas it is also influenced by additional factors such as luminosity, nickel mass, and light curve timescales, creating a tricky interdependence that shapes the resulting bolometric light curves of SNe.\\

    \item \spin{} $-$ \lp{} and $B$ $-$ \lp: A negative correlation of $-$0.72 and $-$0.6 respectively between \spin{} - \lp{} and $B$ - \lp{} suggest that supernovae powered by slower-spinning (higher \spin) and highly magnetised magnetars tend to have the lower peak luminosities. However, this is not unexpected and follows directly from the relationship $P_i \propto E_p^{-0.5}$, where \ep{} represents the magnetar's initial rotational energy. In the magnetar model, the supernova light curve is driven by the deposition of this rotational energy into the supernova ejecta. Therefore, \lp{} is expected to scale with \ep. Since \spin{} is inversely related to \ep, this implies that \lp{} will also be inversely proportional to \spin. Similarly, as discussed, the positive correlation between $B$ and \spin{} are positively correlated which results in a negative correlation between $B$ and \lp, see also \cite{Metzger2015} for \spin{} and $B$ dependence on \lp{} along with other parameters. See the lower middle and right panels of Figure~\ref{fig:scatter_plot_pairs} for the scatter plots of \spin{} vs. \lp{} and $B$ vs. \lp, respectively, which appear to exhibit reciprocal behaviours.

\end{itemize}

\subsection{Dimensionality Reduction}

\begin{figure}
\centering
\includegraphics[angle=0,scale=0.5]{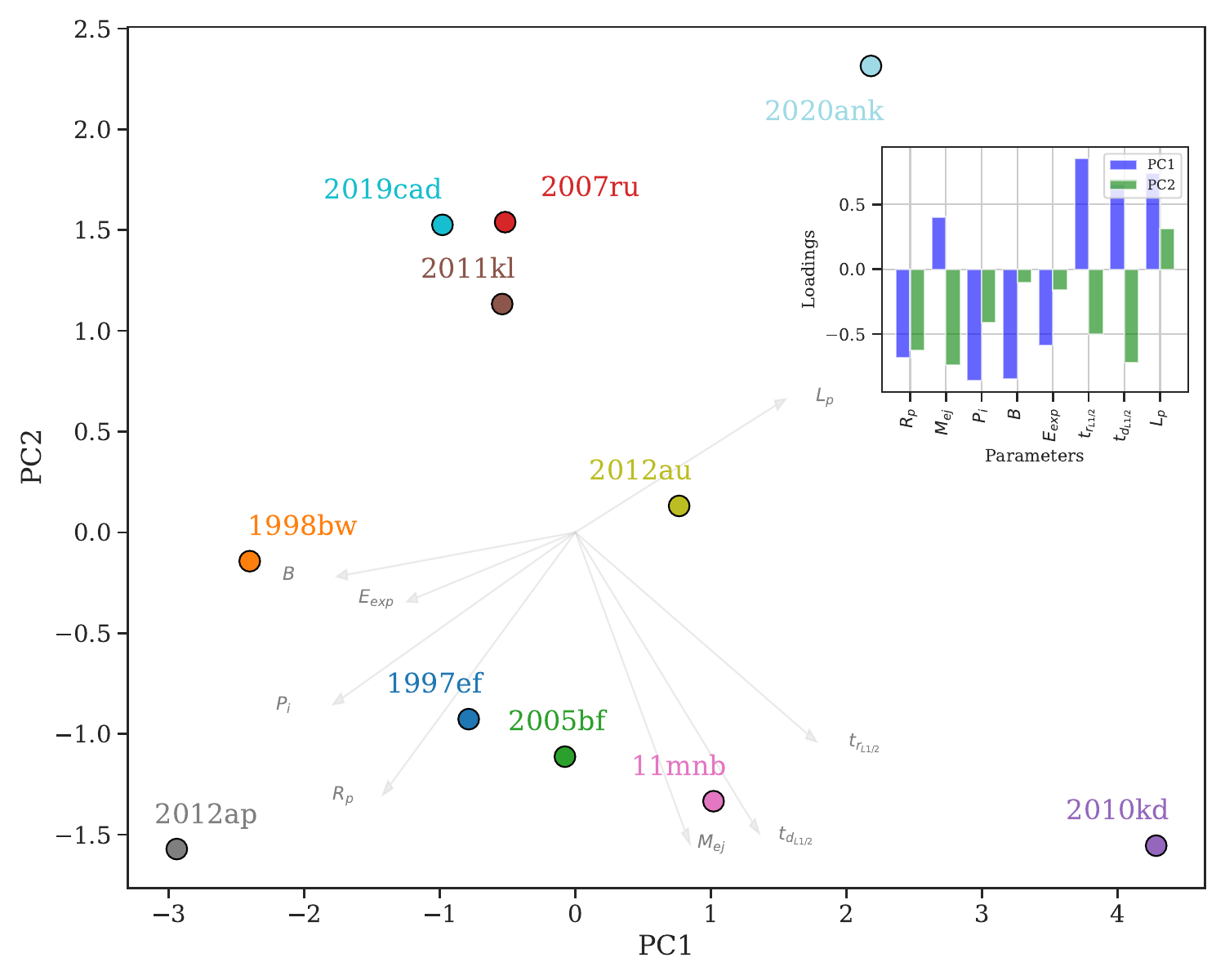}
\caption{PCA reveals that $\sim$70\% of the variance in the dataset is captured by the first two principal components (PC1 - 47\% and PC2 - 23\%), whereas the first four components collectively explain about 90.7\% of the variance in the data, which is sufficient for visualizing and interpreting key trends. The directions of the original 8 parameters used for the PCA (arrows in grey colours) in PC1 - PC2 space and the inset plot with Loading vs. Parameters in the upper-right corner highlight the contributions of each parameter to PC1 and PC2.}
\label{fig:PCA}
\end{figure}

To further investigate the properties of SESNe based on the parameters derived from the light curve analysis, Principal Component Analysis (PCA, \citealt{Pearson1901}) is applied to the set of parameters used in the correlation analysis in Section.~\ref{sec:correlation}. PCA reduces the dimensionality of the dataset by transforming a multi-dimensional parameter space into orthogonal principal components (PCs), which are linear combinations of the original parameters. The PCs capture the maximum variance in the dataset while minimizing redundancy, reducing any overlap in the information shared by each component. By identifying the main directions in which the data exhibits the most significant variation, PCA achieves effective dimensionality reduction, retaining key patterns and relationships within the data. This reduction makes it easier to analyze complex datasets efficiently, maintaining essential features. In the current work, PCA enables clearer visualization and more interpretable insights into different types of SESNe in multi-dimensional parameter space (8 parameters: \rp, \mej, \spin, $B$, \eexp, \trhalf, \tdhalf{} and \lp{}; corresponds to 8-dimensional parameter space) but with a lower-dimensional, simplified representation (2-dimensional space: PC1 and PC2). The PCA results indicate that the first two principal components (PC1 and PC2) capture approximately 70\% of the dataset’s variance, with PC1 contributing 47\% and PC2 adding 23\%. Together, the first four components explain around 90.7\% of the variance, providing a solid foundation for visualizing and interpreting the main trends in the data. The positions of 11 SESNe in PC1 - PC2 space are displayed in Figure~\ref{fig:PCA} with circle signs of different colours, and their locations allow for a comparison of parameter distributions, highlighting commonalities and differences among their parameters.

Based on the Loading vs. Parameters inset-plot and the directions of original parameters in PC1 - PC2 space in Figure~\ref{fig:PCA}, PC1 seems to have high contributions (loading $>$ +0.7 or $< -$0.7) from $B$ and \spin{} in the negative direction and from \trhalf{} and \lp{} parameters in the positive direction. Whereas, PC1 has a moderate contribution (loading +0.4 to 0.7 or $-$0.4 to $-$0.7) from \eexp{} in the negative direction and from \tdhalf{} in the positive direction. On the other hand, PC2 contains high contributions from \mej{} and \tdhalf, and moderate contribution from the \trhalf{} in the negative direction; whereas none of the parameters has a significant contribution in the positive direction. Furthermore, \rp{} moderately but equally contributes to the PC1 and PC2 in the negative direction, hence considered neutral (see the inset plot in the upper-right corner of Figure~\ref{fig:PCA}). Parameters $B$, \spin, \eexp{} and \rp; and \mej, \trhalf{} and \tdhalf{} respectively show close clusters, indicating their possible close correlation, whereas \lp{} lies in the negative direction of \spin{} suggesting their negative correlation, which can also be seen in the correlation analysis presented in Section~\ref{sec:correlation}.

Among 11 SESNe in PC1 - PC2 plane in Figure~\ref{fig:PCA}, SNe 2007ru, 2011kl and 2019cad; and SNe 1997ef, 2005bf and PTF11mnb form relatively compact groups, suggesting that their physical parameters share commonalities. SNe 2007ru, 2011kl and 2019cad share closer values of \mej, \trhalf{} and \tdhalf; and SNe 1997ef, 2005bf, PTF11mnb present comparatively closer values of \rp, \eexp{} and \mej. On the other hand, SNe 1998bw, 2012ap, 2012au, 2020ank and 2010kd are comparatively isolated and share different locations on the PC1 - PC2 plane, suggesting that these SNe exhibit distinct characteristics. SNe 1998bw and 2012ap share the highest values of \rp{} and \eexp, but their different positions on the PC1 - PC2 plane could arise because of huge differences among their $B$ values. SN 2012au holds a nearly central location on the PC1 - PC2 plane, indicating its balance between different parameters and it seems like an intermediate event among clustered SNe. The locations of both SLSNe 2010kd and 2020ank are well isolated in comparison to other SNe and each other; their lowest values of \spin{} and $B$ place them more isolated from other SESNe in the group, whereas, the highest values of \mej, \trhalf{} and \tdhalf{} of 2010kd than other SESNe and also than those of 2020ank place it isolated than other SESNe and also SLSN 2020ank. Overall, within this limited sample, no clear clustering is observed among SNe of the same SESN subtypes (e.g., Type Ic and SLSNe-I), suggesting that similar SESN types can exhibit distinct physical parameter values. These deviations underscore the complexity and diversity within the population of SESNe, providing key insights into the diverse nature of their powering sources, progenitors and explosion dynamics. However, a study with a more extensive sample is needed to confirm any commonalities or discrepancies among their physical parameters.

\section{Summary}
\label{sec:summary}

This work focuses on semi-analytical light curve modelling for a sample of 11 SESNe, considering that these light curves are driven by spin-down millisecond magnetars. The sample includes a diverse set of SESNe subtypes: Type Ib SN 2012u, Ibc SN 2005bf, Ic SNe 1997ef, 2007ru, PTF11mnb, and 2019cad, the relativistic Ic-BL SN 2012ap, GRB-SN 1998bw, GRB-SLSN 2011kl, and superluminous SNe 2010kd and 2020ank. The SNe in this sample span a wide range of peak bolometric luminosities, from approximately $2 \times 10^{42}$ to $4 \times 10^{44}$ erg s$^{-1}$, with SN 1997ef showing the lowest and SN 2020ank the highest peak luminosity.

The MAG model under the {\tt MINIM} code well-regenerates the light curves of all the SESNe in the sample, achieving $\chi^2$/dof values close to one, except for PTF11mnb, which shows significant data scatter. This light curve modelling allows constraining various physical parameters for SNe in the sample, e.g., \ep, \td, \tp, \rp, \vexp, \mej, \spin, $B$ and \eexp. In addition, \lp, \trhalf{} and \tdhalf{} are estimated using the GP regression fitting on bolometric light curves of SESNe in the sample.

As anticipated, SLSNe and Ib/c SNe in the sample exhibit lower \vexp{} ($\approx$12,000 km s$^{-1}$) than those of Ic-BL SNe ($\gtrsim$20,000 km s$^{-1}$). SNe Ib/c in the sample (except SN 1997ef) share comparable \spin{} values around 20 ms, whereas SLSNe 2010kd and 2020ank exhibit the lowest \spin{} around 2.3 ms. On the other hand, SLSNe 2010kd and 2020ank display the lowest and relativistic Ic-BL SN 2012ap shows the highest values of $B$ and \spin values. All the SESNe in the sample (except SN 2019cad) exhibit \eexp{} $>$2 $\times$ 10$^{51}$ erg, suggesting JJEM could be their possible explosion mechanism. 

This work further investigates Pearson’s correlations among key physical parameters of SESNe derived from light curve analysis. The correlation analysis indicates strong positive correlations between \trhalf{} and \tdhalf, \spin{} and $B$, \spin{} and \rp, and \eexp{} and \rp, as well as strong anti-correlations of \spin{} and $B$ with \lp. Additionally, the study examines the distribution of SESNe in the PC1 - PC2 space, obtained through PCA-based dimensionality reduction of the multidimensional parameter space of physical parameters of SESNe. The PC1 - PC2 plane from PCA reveals two compact clusters --- SNe 2007ru, 2011kl, and 2019cad; and SNe 1997ef, 2005bf, and PTF11mnb --- that occupy similar regions in parameter space. While SNe 1998bw, 2012ap, 2012au, 2020ank, and 2010kd appear more isolated, indicating distinct characteristics. This study highlights the unique physical behaviours across SESNe and highlights the value of further exploration of SESNe diversity with an extended sample.

\section*{Data Availability}

The publicly available data used in this paper can be accessed through the respective references cited. Additional data supporting this study are available upon request from the corresponding author.

\section*{Acknowledgments}
AK is supported by the UK Science and Technology Facilities Council (STFC) Consolidated grant ST/V000853/1. AK is thankful to Dr. Claudia Guti{\'e}rrez for sharing the data. AK expresses gratitude to Prof. Jozsef Vink{\'o} and Dr. Kaushal Sharma for providing invaluable insights related to this work. I sincerely thank the anonymous referee for their constructive comments that improved this work. I also acknowledge the invaluable support of NASA's Astrophysics Data System Bibliographic Services.

\bibliography{manuscript.bib}

\end{document}